\documentclass[12pt]{iopart}

\usepackage{iopams}

\date{06/07/2006}

\newcommand{\be}{\begin{eqnarray}}
\newcommand{\ee}{\end{eqnarray}}

\begin{document}
\title[DBI]{Geometrical dynamics of Born-Infeld objects}
\author{Ruben Cordero\dag, Alberto Molgado\ddag\ and Efrain Rojas\S}
\address{\dag\ Departamento de F\'\i sica, Escuela Superior de 
F\'\i sica y Matem\'aticas del I.P.N., 
Unidad Adolfo L\'opez Mateos,
Edificio 9, 07738 M\'exico, D.F., MEXICO }
\address{\ddag\ Facultad de Ciencias, Universidad de Colima,
Bernal D\'\i az del Castillo 340, Col. Villas San Sebasti\'an,
Colima, MEXICO }
\address{\S\ Facultad de F\'\i sica e Inteligencia Artificial,
Universidad Veracruzana, 91000 Xalapa, Veracruz, MEXICO}

\eads{
\mailto{cordero@esfm.ipn.mx}, \mailto{albertom@ucol.mx},
\mailto{efrojas@uv.mx}}

\begin{abstract}
We present a geometrical inspired study of the
dynamics of  $Dp$-branes. We focus on the usual nonpolynomial
Dirac-Born-Infeld action for the worldvolume swept out by the
brane in its evolution in general background spacetimes.
We emphasize the form of the resulting equations of motion 
which are quite simple and resemble Newton's second law, 
complemented with a conservation law for a worldvolume bicurrent.
We take a closer look at the classical Hamiltonian analysis
which is supported by the ADM framework of general relativity.
The constraints and their algebra are identified as well as the 
geometrical role they play in phase space.  In order to 
illustrate our results, we 
review the dynamics of a 
$D1$-brane immersed in a $AdS_3 \times S^3$ background spacetime. 
We exhibit the mechanical properties of 
Born-Infeld objects paving the way to a consistent quantum 
formulation.
\end{abstract}

\submitto{\CQG}


%\maketitle

\section{Introduction}

Nowadays $M$/string theory still is the best candidate to unify all
fundamental interactions. Its non-perturbative approach has
revealed certain important higher dimensional extended objects
known as $Dp$-branes. These objects are defined as hypersurfaces 
in spacetime onto which open strings can attach, and Dirichlet
boundary conditions are chosen for them~\cite{DLP, Horava, Polchinski1}.
Since its discovery, a lot of effort has
been devoted to the study of $Dp$-branes due to the key role 
they have played in the understanding of physics at the tiniest scales.
$Dp$-branes are enlightening when non-perturbative properties
of superstring theory and $M$-theory are studied~\cite{Polchinski1}. They are
also relevant in the quantum description of black holes~\cite{CM, SV} 
and the geometric nature of
spacetime~\cite{Clifford}.  Besides,  
they bring insights into new scenarios for cosmological theories 
of our entire universe~\cite{Arkani,Antoniadis,RS}.  There are, 
in addition, some other remarkable contributions in high-energy 
physics provided by $Dp$-branes (see~\cite{Clifford, Polchinski2}, 
and references therein).

The Dirac-Born-Infeld (DBI) action has been proposed as
an elegant effective action governing the dynamics of $Dp$-branes at 
low energy scales \cite{Polchinski1,Clifford}. Originally, 
Born-Infeld (BI) theory arose to overcome the infinity problem 
associated with the self-energy of a point charge source in the 
classical Maxwell theory \cite{Born,Born1}. The simplest DBI 
Lagrangian in the realm of string theory is based on the introduction 
of a $U(1)$ gauge field propagating on the worldvolume swept out 
by an extended object coupled to the inherent geometry of the 
worldvolume. It is one of the simplest nonpolynomial Lagrangians 
invariant under reparametrizations of the worldvolume. In fact, 
the DBI action is the natural generalization of the Dirac-Nambu-Goto 
(DNG) action which describes minimal surfaces but now involving 
a gauge field on the worldvolume. 

In this paper we aim to perform a Lagrangian and a Hamiltonian 
geometrical study of the DBI action. We obtain a geometrical
interpretation of the mechanical properties of $Dp$-branes which
constitutes the mathematical backbone of the paper. The variational
process of the DBI action leads to second order equations of motion
that are complemented with a conservation law associated to the energy-momentum,
on one side, and Maxwell's equations
arising from the conservation of a bicurrent defined on the worldvolume, on the other.
The equations of motion associated to the coordinates are simply the
contraction of the worldvolume stress tensor with the extrinsic curvature
equated to an external force. This can be interpreted as a generalized
Newton's second law. The corresponding conserved momentum is 
constructed with two terms: the kinetic momentum and the 
interaction of the $Dp$-brane with a Neveu-Schwarz (NS) field. 
This result generalizes the conserved momentum for a point particle interacting with an electromagnetic field.
As we will see, the source of the
currents and external forces mentioned above 
resides in the presence of the antisymmetric 
NS field.
We presume that the geometrical language
is more convenient for the general covariant analysis of $Dp$-branes.

Even though there are good studies on the DBI action we are convinced 
that some special issues need a careful analysis. We feel unpleasant 
with other approaches where sometimes the simplification of the original 
action by means of auxiliary variables leads to an associated murky 
geometrical content. These alternative ways to study the DBI action are
based on the introduction of nondynamical auxiliary fields where the resulting 
action is nondeterminantal and adquires a linear dependence on the 
derivatives of the fields. The obtained simplification by means of this 
method is useful for certain type of calculations but nevertheless it 
conceals the study of inherent geometric properties of the theory. Excellent 
studies dealing with this approach exist in the literature 
(see~\cite{Zeid, Lee, Bozhilov}, 
for example).
Furthermore, at the Hamiltonian level, these approaches start by considering 
an equivalent DBI action where square root terms are avoided in 
the Lagrangian, leading to a lenghty canonical formulation 
where a plethora of constraints emerge,  
hiding the geometrical structure of the system and making mechanical 
interpretation unclear.

A different approach, which we will follow here, is to consider the original
square root DBI Lagrangian density and to derive the physical content
by appealing basic notions in differential geometry. Our approach does not
confront other existent Hamiltonian approaches for the DBI action but, on the
contrary, it clarifies more the geometrical role of the constraints and
elucidates the geometrical nature of the momenta. 
%This analysis is well suited for newcomers in BI-like objects. 
Though for simplicity we shall restrict 
our description to $U(1)$ gauge fields living on the worldvolume, we believe 
that the inclusion of other fields, does not change considerably the structure 
of our approach. 
Moreover, since our approach consider general backgrounds, we believe 
that our geometrical Hamiltonian analysis is useful to study the dynamics of 
other physically interesting objects like supertubes, for example~\cite{kluson2,supertubes}. 
For the sake of 
simplicity we assume in this paper that the $Dp$-branes have not spatial 
boundaries or that the physical fields fall off appropriately at large distances. 

The rest of the paper is organized as follows. In
section~\ref{sec:DBIaction} we write the DBI action and we obtain the
equations of motion in terms of the worldvolume stress tensor,
a covariant worldvolume bicurrent and conserved quantities. 
In section~\ref{sec:ADM} we take advantage of the ADM formulation
of General Relativity (GR) in order to decompose the trajectory of the
$Dp$-brane in a similar fashion. Section~\ref{sec:constraints} is devoted
to develop the Hamiltonian analysis of the theory identifying the
constraints and their role in phase space. We also find the
algebra of the constraints. In section~\ref{sec:Hamilton} we write
Hamilton's equations of motion in terms of the geometrical
quantities described before. 
We put our machinery at work  
by considering an unambiguous example in section~\ref{sec:kluson}
in an anti-deSitter background previously analysed using a different approach 
in~\cite{Bachas, KNP}. 
Section~\ref{sec:remarks}
presents some concluding remarks and some perspectives on the quantum
approach.
Finally, in~\ref{app:examples} we specialize our results 
to the $D1$-brane and $D2$-brane general cases, explicitly showing the 
form of the worldvolume conserved quantities, and 
\ref{app:mathidentities} collects some mathematical identities 
used in the main text.  

\section{DBI action}
\label{sec:DBIaction}

Consider a $Dp$-brane, $\Sigma$, of dimension $p$ evolving in 
a $N$-dimensional background spacetime endowed with an arbitrary  
metric $G_{\mu \nu}$ $(\mu, \nu = 0,1,2, \ldots, N-1)$. The 
trajectory, or worldvolume, $m$, swept out by $\Sigma$ is an 
oriented timelike manifold of dimension $p+1$, described by the 
embedding functions $x^\mu = X^\mu(\xi^a)$ where $x^\mu$ are 
local coordinates of the background spacetime, $\xi^a$ are 
local coordinates of $m$, and $X^\mu$ are the embedding functions 
$(a,b= 0,1,2,\ldots,p)$. The metric induced on the worldvolume 
from the background is given by $g_{ab}= G_{\mu \nu}X^\mu
{}_a X^{\nu}{}_b := X_a \cdot X_b$ with $X^\mu{}_a=\partial_a X^\mu=
\partial X^\mu /\partial \xi^a$ being tangent vectors to $m$.

In this framework we introduce $N-p-1$ normal vectors to the 
worldvolume, denoted by $n^\mu {}_i \,\,(i=1,2,\ldots,N-p-1)$. 
These are defined implicitly by $n\cdot X_a = 0$ and we choose 
to normalize them as $n_i \cdot n_j = \delta_{ij}$. We will adopt 
index-free notation when convenient in order to avoid a cumbersome 
description.

The celebrated effective nonpolynomial DBI action that controls 
the low energy dynamics of $Dp$-branes is
\begin{equation}
S_{{\mbox{\tiny DBI}}}[X^\mu,A_a]=  \beta_p \int_m d^{p+1}\xi \,
\sqrt{-{\mbox{det}} (g_{ab} + {\cal F}_{ab})}\,,
\label{eq:DBIaction}
\end{equation}
where $\beta_p$ is the tension of the $Dp$-brane\footnote[1]{The
explicit form of the $Dp$-brane tension is given by $\beta_p =2\pi
/[g_f (2\pi\sqrt{\alpha'})^{p+1}]$, where $\alpha'$ is the inverse of the
fundamental string tension and $g_f$ is the string coupling.},
${\cal F}_{ab}=\alpha F_{ab} + 
B_{ab}$ with $F_{ab} =2\partial_{[a}A_{b]}$ 
being the electromagnetic field strength associated to a $U(1)$ gauge 
field $A_a$ living on $m$ and $B_{ab} = B_{\mu \nu}X^\mu {}_a X^\nu{}_b$
is the pullback to the worldvolume of the NS 2-form 
$B_{\mu \nu}$; here $\alpha$ is the BI parameter related to the 
inverse tension of fundamental strings\footnote[2]{To enrich the 
content of the work, the parameter $\alpha=2\pi \alpha'$ can be 
thought of as a parameter included in order to gain control over 
the theory.}. The configuration space ${\cal C}$ is spanned by 
$\left\lbrace X^\mu, A_a \right\rbrace$, that is, we have $N + p 
+ 1$ configuration degrees of freedom per point on the worldvolume. 
We are restricting ourselves to consider gauge fields $A_a$ living 
on the worldvolume only. Different BI-like theories exist that 
consider the housing of other gauge fields \cite{search}. In particular, 
the DBI action (\ref{eq:DBIaction}) is invariant under worldvolume 
reparametrizations.\footnote{Some consequences of this reparametrization 
invariance are studied in~\cite{Fairlie}, where quadratic terms under the 
square root of the Lagrangian are studied.} 
Further, the action is invariant under a NS gauge 
transformation $B_{ab} \to B_{ab} - 2\partial_{[a} \lambda_{b]}$ if 
we shift the $U(1)$ field $A_a \to A_a + \alpha^{-1} \lambda_a$ 
where $\lambda_a$ is a $1-$form; in other words, ${\cal F}_{ab}$ is 
the gauge invariant quantity in the presence of NS background field
and not the electromagnetic field tensor $F_{ab}$. For the sake of 
simplicity throughout the paper we will introduce the following 
notation: $M_{ab}:= g_{ab} + {\cal F}_{ab}$ is the composite matrix, 
while $(M^{-1})^{ab}$ denotes its inverse and $M:= {\mbox{det} 
(M_{ab})}$. Henceforth, as a further notational simplification, we 
will omit the differential symbols, $d^{p+1}\xi$ or $d^p u$, wherever 
a worldvolume or space integration is performed.

Under an infinitesimal deformation of the embedding functions
$X \to X + \delta X$ as well as $A \to A + \delta A$, the first
variation of the action (\ref{eq:DBIaction}) casts out the
equations of motion associated to the configuration space
${\cal C} $~\cite{KNP,defo},
\begin{eqnarray}
-\nabla_a \left\lbrace \beta_p \frac{\sqrt{-M}}{\sqrt{-g}} \left[ 
(M^{-1})^{(ab)} G_{\mu \nu} -  (M^{-1})^{[ab]} B_{\mu \nu} 
\right] \,X^\nu {}_b
\right\rbrace & &\nonumber \\
+ \frac{\beta_p}{2}\frac{\sqrt{-M}}{\sqrt{-g}} \left[ 
(M^{-1})^{(ab)} \partial_\mu G_{\alpha \nu} -  (M^{-1})^{[ab]} 
\partial_\mu B_{\alpha \nu} \right] \,X^\alpha {}_a X^\nu {}_b
&=& 0\,,
\label{eq:moto1}
\\
\hspace{29ex}
\nabla_a \left[ -\alpha \beta_p\,\frac{\sqrt{-M}}{\sqrt{-g}}
(M^{-1})^{[ab]}\right]
&=& 0\,,
\label{eq:moto2}
\end{eqnarray}
where $\nabla_a$ is the covariant derivative associated with
$g_{ab}$ and $(M^{-1})^{(ab)}$ and $(M^{-1})^{[ab]}$ denote the 
symmetric and the antisymmetric parts of the matrix $(M^{-1})$, 
respectively. Note that we have $N + p + 1$ equations of motion, some
of them being simple identities whose origin is related to the
invariance under reparametrizations of the worldvolume.

It is convenient to compute the worldvolume stress-energy
tensor defined by
$T^{ab}:= \frac{2}{\sqrt{-g}}
\frac{\delta S_{{\mbox{\tiny DBI}}}}{\delta g_{ab}}
$,
and the worldvolume covariant bicurrent
defined by ${J}^{ab}:=\frac{2}{\sqrt{-g}} 
\frac{\delta S_{{\mbox{\tiny DBI}}}}{\delta F_{ab}}$ 
(see~ \cite{Gibb}). We obtain the general expressions
\begin{eqnarray}
T^{ab} &=& \beta_p \,\frac{\sqrt{-M}}{\sqrt{-g}}
(M^{-1})^{(ab)}\,,
\label{eq:Tab}
\\
{J}^{ab}&=& -\alpha \beta_p\,\frac{\sqrt{-M}}{\sqrt{-g}}
(M^{-1})^{[ab]}\,.
\label{eq:Jab}
\end{eqnarray}
Despite the definition of the worldvolume bicurrent,
(\ref{eq:Jab}) establishes a nonlinear relation between 
${J}^{ab}$ and $F^{ab}$ as we shall see below. In fact,
$(\ref{eq:Jab})$ denotes the covariant form of the electric
induction on the worldvolume \cite{Gibb}. A well known relation 
exists between the physical tensors $T^{ab}$ and
$J^{ab}$ which was introduced since the foundation
of the BI theory. It can be obtained easily mixing the expressions 
defining the Eqs.~(\ref{eq:Tab}) and (\ref{eq:Jab}) as well as the
identity (\ref{eq:ident1}),
\begin{equation}
 T^a{}_c = (-g)^{-1/2}{\cal L}\,\delta^a{}_c - \alpha^{-1}
J^{ba}{\cal F}_{bc}\,.
\label{eq:first-identity}
\end{equation}
Taking into account (\ref{eq:Tab}) and (\ref{eq:Jab}) we can
rewrite the equations of motion~(\ref{eq:moto1}) and 
(\ref{eq:moto2}) in terms of the tensors $T^{ab}$ and 
$J^{ab}$ as
\begin{eqnarray}
\fl
 \nabla_a \left[ \left( T^{ab} G_{\mu \nu} + \alpha^{-1}J^{ab} 
B_{\mu \nu} \right) X^\nu {}_b
\right] - \frac{1}{2}  \left( T^{ab} \partial_\mu G_{\alpha \nu} +
\alpha^{-1}J^{ab} \partial_\mu B_{\alpha \nu} \right) 
X^\alpha {}_a X^\nu {}_b
&=& 0\,,
\label{eq:Moto1}
\\
\hspace{52.5ex}
\nabla_a J^{ab} &=& 0\,.
\label{eq:Moto2}
\end{eqnarray}

From the relation (\ref{eq:first-identity}) and with
the help of the conservation of the bicurrent $J^{ab}$, it is
straightforward to show that the metric stress tensor $T^{ab}$ is also
conserved, 
\begin{equation}
 \nabla_a T^{ab} = 0\,.
\label{eq:eqmoto0}
\end{equation}
Now, based on this result and with the help of the Gauss-Weingarten equations
$\nabla_a X^\mu_b = -\Gamma^\mu _{\alpha \beta} X^\alpha{}_a X^\beta{}_b
- K_{ab} ^i n^\mu {}_i$, where $ \Gamma^\mu _{\alpha \beta}$ are the Christoffel
symbols associated to $G_{\mu \nu}$, the equations of motion~(\ref{eq:Moto1}) 
take the form
\begin{equation}
 -T^{ab} K_{ab} ^i G_{\mu \nu} n^\nu {}_i + \alpha^{-1} \frac{1}{2}
J^{ab} (\partial_\alpha B_{\mu \nu} + \partial_\nu B_{\alpha \mu} 
+ \partial_\mu B_{\nu \alpha}) X^\alpha {}_a X^\nu {}_b = 0\,,
\end{equation}
where $K_{ab} ^i= - n^i \cdot D_a X_b$ is the extrinsic curvature of the worldvolume 
and $D_a = X^\mu {}_a D_\mu$ is the pullback to the worldvolume of the covariant 
derivative compatible with $G_{\mu \nu}$, that is, 
$D_\mu$~\cite{defo}. Finally, the equations of motion adquiere the geometrical 
simplified form
\begin{eqnarray}
 T^{ab} K_{ab} ^i&=& F^i\,,
\label{eq:eqmoto1}
\\
\nabla_a J^{ab} &=& 0\,,
\label{eq:eqmoto2}
\end{eqnarray}
where $F^i= - \frac{1}{2}\alpha^{-1} J^{ab} H_{\alpha \beta \mu}
X^\alpha {}_a X^\beta {}_b n^{\mu\,i}$ with $H_{\alpha \beta \mu}
= \partial_\alpha B_{\mu \nu} + \partial_\nu B_{\alpha \mu} 
+ \partial_\mu B_{\nu \alpha}$ being  the NS strength 3-form field
which satisfies $\partial_\mu H^{\mu \alpha \beta}=0$.
Note that the form of the equations of motion (\ref{eq:eqmoto1}) can 
be interpreted as a generalization of Newton's second law for a 
particle where $T^{ab}$ plays the role of a mass, $K_{ab} ^i$ the 
generalization of the acceleration in higher dimensions, and $F^i$ 
a force density. This form of the equations of motion was obtained 
in other contexts, for instance, in the case of superconducting 
membranes and membranes interacting with external Kalb-Ramond and 
$U(1)$ fields in~\cite{Carter1, RE,Carter2}.

The classical trajectories of $Dp$-branes are governed by $N - 
p -1$ independent equations of motion (\ref{eq:eqmoto1}) of second 
order in the coordinates $X$, one for each normal. The $(p+1)$ 
equations of motion (\ref{eq:eqmoto2}) are associated to the $U(1)$ 
gauge fields $A_a$. The remaining $(p+1)$ tangential equations 
(\ref{eq:eqmoto0}) are satisfied identically as a consequence of the
reparametrization invariance of the action $S_{DBI}$. 
As expected, the DNG equations of motion $K^i = g^{ab}K_{ab}^i=0$ in the 
context of minimal surfaces, are recovered when we turn off the fields. We 
reassert that the equations of motion are a generalization of the minimal 
surface equations with the addition of an $U(1)$ gauge field living on the 
worldvolume. With respect to the equations (\ref{eq:eqmoto2}), these yield 
Maxwell equations either homogeneous or inhomogeneous with support on the 
worldvolume. Note that with this simple derivation of the dynamics, we have 
a proof of the conservation of both the stress-energy tensor $T^{ab}$ and 
the bicurrent ${J}^{ab}$. To make our approach more concrete  
in~\ref{app:examples} we specialize to the $D1$- and $D2$-brane general 
cases, where we explicitly exhibit the worldvolume conserved quantities. 
We expect that these cases provide us with enough intuition to understand 
the evolution of $Dp$-branes for general $p$.

With respect to the conserved quantities, we must assume 
that the background spacetime has certain symmetries. In general, 
the response of the action (\ref{eq:DBIaction}) under an 
infinitesimal deformation of the embedding functions $X \to X + 
\delta X$ as well as $A \to A + \delta A$ can be expressed by
\begin{equation}
\delta S_{{\mbox{\tiny DBI}}} = \int_m \sqrt{-g}\left[ {\cal E}_\mu
\delta X^\mu + {\cal E}^a \delta A_a + \nabla_a Q^a \right]\,,
\end{equation}
where ${\cal E}_\mu$ and ${\cal E}^a$ are the Euler-Lagrange
derivatives of the action and $Q^a$ is the Noether charge that
depends on the infinitesimal deformations $\delta X$ and $\delta
A$ (see \cite{Noether} for details). For an infinitesimal constant
deformation $\delta X^{\mu} = \epsilon^{\mu}$, and assuming that 
the equations of motion are satisfied, the variation of the action 
becomes
\begin{equation}
\delta S_{{\mbox{\tiny DBI}}} = \epsilon^{\mu} \int_m \sqrt{-g}\nabla_a Q^a _{\mu}\,,
\label{Noether}
\end{equation}
where we can set the Noether charge as
\begin{equation} 
Q^a _{\mu} = T^{ab}\,X^\nu {}_{b}G_{\mu\nu} + \frac{1}{\alpha}J^{ab}B_{\mu\nu} X^{\nu}{}_{b}\,.
\end{equation}
By using the divergence theorem, equation~(\ref{Noether}) reads
\begin{equation}
\delta S_{{\mbox{\tiny DBI}}} = \epsilon^{\mu}\int_{\partial m}\sqrt{h}
\eta_a Q^a _{\mu} = \epsilon^{\mu}\int_{\Sigma_{f}}\sqrt{h}\eta_a Q^a _{\mu} - \epsilon^{\mu}\int_{\Sigma_{i}}\sqrt{h}\eta_a Q^a _{\mu} \, ,
\end{equation}
where the integrals are evaluated on the spacelike hypersurface $\Sigma$ 
at initial and final times.  The spacelike hypersurface $\Sigma$ seen as 
embedded in $m$, has a timelike normal vector $\eta_{a}$ and the determinant 
of its induced metric is $h$ (see below). If the action is invariant under 
$\delta X^{\mu} = \epsilon^{\mu}$ we have the constant of motion (because 
$\Sigma_f$ and $\Sigma_i$ are arbitrary),
\be
{\cal P}_{\mu} = \int_{\Sigma}\sqrt{h}\eta_a (T^{ab}\,X^\nu {}_{b}G_{\mu\nu} + \frac{1}{\alpha}J^{ab}B_{\mu\nu} X^{\nu}{}_{b})\, .
\ee
This result is pretty nice since it generalizes the conserved momentum for 
a relativistic particle interacting with an electromagnetic field. In our 
case the first term represents the kinetic momentum and the second one is 
the coupling between the $Dp$-brane and the NS field.

If we have a static background, the action is invariant under time 
translations and the corresponding conserved quantity will be the energy 
of the $Dp$-brane moving in the background spacetime
\begin{equation}
{\cal P}_{0} = \int_{\Sigma}\sqrt{h}\eta_a (T^{ab}\,X^\nu {}_{b}G_{0\nu} + \frac{1}{\alpha}J^{ab}B_{0\nu} X^{\nu}{}_{b})\, .
\end{equation}
One can further note immediately from equation~(\ref{eq:Moto1}) that in the case of a
static background the energy density $\varepsilon_p$ of a
$Dp$-brane is conserved, $\partial_t \varepsilon_p=0$, where
\begin{equation}
 \varepsilon_p = \sqrt{-g}(T^{ta}G_{0\mu} + \alpha^{-1}J^{ta}B_{0\mu})X^\mu{}_a\,.
\label{eq:energy}
\end{equation}
In a similar fashion, for a gauge transformation $\delta A = \partial \Lambda$ the 
corresponding Noether charge is
\begin{equation}
\Pi^a = \sqrt{-g}J^{ab} \partial_b \Lambda\,.
\end{equation}
The condition $\delta S_{{\mbox{\tiny DBI}}} =0$ 
for an action invariant under translations as well as
gauge transformations of the kind $\delta A = \partial \Lambda$ 
imposes the Euler-Lagrange
derivatives as
\begin{equation}
{\cal E}^\mu = - \nabla_a Q^{a\, \mu}\,,
\end{equation}
and
\begin{equation}
{\cal E} = - \nabla_a \Pi^a = \sqrt{-g}{\cal E}^a \partial_a \Lambda\,.
\end{equation}
In other words, the equations of motion can be restated in terms of the 
conservation of the tensors $Q^{a\, \mu}$ and $\Pi^a$.

\section{ADM decomposition}
\label{sec:ADM}

We shall consider that the worldvolume $m$ is generated by the evolution of
$\Sigma$ in a parameter $t$. The $Dp$-brane $\Sigma$ is described locally
by the embedding $x^\mu = X^\mu (t,u^A)$ at fixed
$t$, where $u^A$ are local coordinates on $\Sigma$,
$(A,B=1,2,\ldots,p)$. The parameter $t$ keeps track
of the evolution of $\Sigma$. Imitating the ADM procedure
of GR~\cite{ADM}, the flow of time throughout the worldvolume $m$
is represented  by means of a time vector field
$\dot{X}^\mu =\partial_t X^\mu$.  This is a tangent vector to $m$ which
is expanded conveniently with respect to the basis
$\left\lbrace \eta^\mu, X^\mu {}_A \right\rbrace$ of tangent
vectors to the worldvolume living in $\Sigma$, that is,
\begin{equation}
\dot{X} = N \eta + N^A X{}_A\,,
\label{eq:dotX}
\end{equation}
where $N$ and $N^A$ are denoted as the lapse function and the shift vector
respectively, as in GR. In this decomposition $X^\mu{}_A =
\partial_A X^\mu= \partial X^\mu /\partial u^A$ are the $p$
tangent vectors to $\Sigma$ and $\eta^\mu$ stands for a unit timelike
normal vector to $\Sigma$.

Similarly as in section~\ref{sec:DBIaction}, we can consider
the embedding of $\Sigma$ in the worldvolume $m$, by means of
$\xi^a = X^a (u^A)$, where $X^a$ are embedding functions also.
The metric induced on $\Sigma$ is $h_{AB}= g_{ab}
X^a{}_A X^b{}_B = G_{\mu \nu}X^\mu{}_A X^\nu{}_B$ with $X^a{}_A$
being tangent vectors to $\Sigma$ at fixed $t$, and $\eta^a$ is a
timelike unit vector to $\Sigma$.
The worldvolume vectors $\left\lbrace
\eta^a, X^a {}_A \right\rbrace $ form a basis\footnote{This basis
satisfies the completeness relation
\be
h^{AB}X^a{}_A X^b{}_B =g^{ab}+\eta^a \eta^b= g^{ab} + N^{-2}
(\dot{X}^a - N^A X^a{}_A)(\dot{X}^b - N^BX^b{}_B)\,. \nonumber
\ee
}
for the worldvolume $m$ adapted to $\Sigma$. Both embedddings ($x^\mu
= X^\mu (\xi^a)$ and $\xi^a = X^a (u^A)$ ) are related by composition
\cite{hambranes}.

The idea is to break up the non-symmetric composite matrix,
$M_{ab}:= g_{ab} +  {\cal F}_{ab}$, in its ``temporal" and
``spatial" components. We know how to do it for the case of
$g_{ab}$ by means of the ADM decomposition~\cite{ADM,hambranes}. In
the same spirit, the standard ADM framework of GR
will be adapted to $M_{ab}$. So, first note the following,
\begin{eqnarray}
\dot{X}^a \dot{X}^b {\cal F}_{ab}&=&  0 \,,\\
\dot{X}^a X^b{}_A {\cal F}_{ab} &=& \alpha \left( \dot{A}_A - {\cal D}_A  
A_0 \right)  +B_{0A}:=-\alpha E_A + B_{0A}\,,
\label{eq:A0}
\\
X^a{}_A X^b{}_B {\cal F}_{ab}&=& {\cal F}_{AB} = - {\cal F}_{BA}\,.
\end{eqnarray}
where we have defined $A_0 = A_a \dot{X}^a$ and $A_A = A_a
X^a _A$, and $B_{0A} = \dot{X}^a X^b {}_A B_{ab}$. By ${\cal D}_A$ 
we denote the covariant derivative compatible with $h_{AB}$. 
Of course, at this moment, as in ordinary
Maxwell theory, we could introduce the electric intensity $\bf{E}$
and the magnetic induction ${\bf B}$ by means of $E_A = F_{A0}$ and
$B^A = \frac{1}{2}\epsilon^{ABC}F_{BC}$.

Supported in the coordinate basis $\left\lbrace \dot{X}, X_A
\right\rbrace $, the composite matrix $M_{ab}$ adquires its
ADM form
\begin{equation}
(M_{ab})= \left(
      \begin{array}{cc}
      (-N^2 + N^AN^B h_{AB})&(N^C M_{CB} + {\cal F}_B) \\
      (N^D M_{AD} -  {\cal F}_A)&M_{AB}
      \end{array}
      \right)\,,
\label{eq:Mab}
\end{equation}
where we have introduced the quantity
\begin{equation}
\label{eq:calF}
{\cal F}_A :=  N \eta^a
X^b {}_A {\cal F}_{ab}\,,
\end{equation}
for short in the notation.
The tangent-tangent projection of the composite matrix $M_{ab}$
is given by $M_{AB}= h_{AB} + {\cal F}_{AB}$.  The reciprocal
matrix $(M^{-1})^{ab}$ is explicitly given by
\begin{equation}
\fl
(M^{-1})^{ab} =
\left(
\begin{array}{cc}
\frac{-1}{N^2-{\cal F}_A (M^{-1})^{AB} {\cal F}_B}                              &
\frac{N^B+ (M^{-1})^{CB}{\cal F}_C}{N^2-{\cal F}_A (M^{-1})^{AB} 
{\cal F}_B}    \\
\frac{N^A + (M^{-1})^{CA}{\cal F}_C}{N^2- {\cal F}_A (M^{-1})^{AB} 
{\cal F}_B}     & \frac{(M^{-1})^{AB}-[N^A+ (M^{-1})^{CA}{\cal F}_C]
[N^B+ (M^{-1})^{DB}{\cal F}_D]}{N^2- {\cal F}_A (M^{-1})^{AB} 
{\cal F}_B}
\end{array}
\right)\,.
\label{eq:inverseMab}
\end{equation}
From matrices~(\ref{eq:Mab}) and~(\ref{eq:inverseMab}) we can 
straightforward recover the usual ADM decomposition for DNG 
extended objects by considering a null electromagnetic field 
$(\alpha=0)$  and null NS fields and hence $M_{AB}=h_{AB}$, 
\cite{hambranes,ADMDNG}.

Note that the following identities hold
\begin{equation}
N=-G_{\mu \nu}\eta^\mu\,\dot{X}^\nu \quad {\mbox{and}}\quad
N^A = h^{AB}G_{\mu \nu} \dot{X}^\mu\,X^\nu{}_B\,,
\end{equation}
which shows the explicit dependence of the lapse function and 
the shift vector on the velocity vector (\ref{eq:dotX}), as opposed 
to the situation in GR.

To calculate the determinant appearing in the action
(\ref{eq:DBIaction}), we shall consider the useful identity for
determinants (see~\ref{app:det})
\begin{eqnarray}
M &=& {\mbox{det}}(M_{ab})
   =  {\mbox{det}}(M_{AB})[M_{00} -
      M_{0A}(M^{-1})^{AB}M_{B0}]\,,
\end{eqnarray}
where $(M^{-1})^{AB}$ is the inverse matrix of $M_{AB}$.
We assume the relations $(M^{-1})^{BC}M_{CA} =
M_{AC}(M^{-1})^{CB}=\delta_A {}^B$ to hold.  We also
define the following quantities that will be useful throughout
\begin{eqnarray}
{\cal M} &:=& {\mbox{det}}(M_{AB}) \,,\\
\Delta   &:=&  M_{00} - M_{0A}(M^{-1})^{AB}M_{B0}\,.
\label{eq:det}
\end{eqnarray}
In terms of these quantities the Lagrangian density turns into
\begin{equation}
{\cal L}= \beta_p \sqrt{- {\cal M}\,\Delta}\,.
\label{eq:densityL}
\end{equation}

Taking into account the components of (\ref{eq:Mab}),
the cofactor (\ref{eq:det}) can be written in a more convenient form as
\begin{equation}
\Delta = -N^2 +  (M^{-1})^{AB}{\cal F}_A {\cal F}_B  \,.
\label{eq:Delta2}
\end{equation}
The computation of the conjugated momenta to the configuration
space ${\cal C}$ is now straightforward from the
expressions (\ref{eq:densityL}) and (\ref{eq:Delta2})
\begin{eqnarray}
P_\mu &:=& \frac{\partial {\cal L}}{\partial \dot{X}^\mu}
       =  \frac{\beta_p \, (-{\cal M})}{\sqrt{-M}} \left[ N \eta_\mu
       +  \left( M^{-1}\right)^{[BA]}{\cal F}_A \,X_{\mu \,B}
          \right] + \frac{1}{\alpha} \pi^A B_{\mu \nu} X^\nu {}_A,
\label{eq:P} \\
\pi^A &:=& \frac{\partial {\cal L}}{\partial \dot{A}_A}
       =  \frac{\beta_p\alpha \,(-{\cal M})}{\sqrt{-M}} \left[
          \left(M^{-1}\right)^{(AB)} {\cal F}_B\right] \,,
\label{eq:pi} \\
\pi^0 &:=& \frac{\partial {\cal L}}{\partial \dot{A}_0}
       =  0\,.
\label{eq:pi0}
\end{eqnarray}
The origin of the latter equation is because
there is no term corresponding to time derivative of $A_0$ in the
action. The momenta are densities of weight one as usual
in any field theory. Unlike the Maxwell field theory, in the
DBI case the canonically conjugated momenta $\pi^B$ to the
potential $A_B$ are only proportional to the electric field
$E_A$ through the combination ${\cal F}_A$ defined
in~(\ref{eq:calF}).

\section{Hamiltonian approach}
\label{sec:constraints}

The recipe to get the Hamiltonian description of a constrained
field theory is first to obtain the canonical Hamiltonian by means
of a Legendre transformation with respect to $\dot{X}^\mu$ and
$\dot{A}_a$, hence we have
\begin{equation}
H_0 [X, P;A, \pi] = \int_\Sigma \left( P \cdot \dot{X} + \pi^a
\dot{A}_a \right)  - L [X,\dot{X}, A, \dot{A}]\,,
\end{equation}
where $L[X,\dot{X}, A, \dot{A}]$ is the DBI Lagrangian functional.
We are dealing with an action invariant under worldvolume
reparametrizations and gauge transformations.  As is well known,
at the Hamiltonian level, this implies the existence of
constraints.

\subsection{Constraints}

From the definitions of the canonical momenta, we identify the 
primary constraints by contracting first~(\ref{eq:P}) with $P$ and 
$X{}_A$ and then~(\ref{eq:pi}) with $\pi^A$ and ${\cal F}_{AB}$, obtaining
\begin{eqnarray}
\hspace{-7ex}
C &:= &G^{\mu \nu} \left( P_\mu -\frac{1}{\alpha} B_{\mu \alpha}\pi^\alpha\right) \left( P_\nu - \frac{1}{\alpha} B_{\nu \beta}\pi^\beta \right) + \frac{1}{\alpha^2} \pi^A \pi^B h_{AB} 
+ \beta_p ^2 {\cal M} 
\simeq 0\,,
\label{eq:scalar}
\\
\hspace{-7ex}
C_A &:=& \left(P_{\mu}-\frac{1}{\alpha} B_{\mu \alpha}\pi^\alpha\right)\,X^{\mu}{}_A 
         + \frac{1}{\alpha} \pi^B {\cal F}_{AB} \simeq 0\,,
\label{eq:vector}
\\
\hspace{-7ex}
C_0 &:= &\pi^0 \simeq 0\,,
\label{eq:gauss0}
\end{eqnarray}
where $\pi^\mu := \pi^A X^\mu{}_A$ is the pushforward of the
canonical momenta conjugated to $A_A$.
Constraints (\ref{eq:scalar}) and (\ref{eq:vector}) correspond
to $(p+1)$ diffeomorphisms and (\ref{eq:gauss0}) corresponds
to the $U(1)$ symmetry. In fact, the constraint (\ref{eq:scalar})
generates diffeomorphisms out of $\Sigma$ onto the worldvolume
and (\ref{eq:vector})  generates the diffeomorphisms tangential
to $\Sigma$. We have in number $(p+2)$ primary constraints.
To obtain them we have used the 
identities (\ref{eq:ident1}) and (\ref{eq:ident2}) of~\ref{app:identities}.

Furthermore, the canonical Hamiltonian density vanishes due to
the invariance under reparametrizations of  the worldvolume
\begin{equation}
{\cal H}_0 = P \cdot \dot{X} + \pi^a \dot{A}_a - {\cal L} = 0\,.
\end{equation}

According to the Dirac-Bergmann analysis for constrained 
systems~\cite{Hen}, we need to evolve in time the phase space 
functions of the theory by means of a total Hamiltonian 
constructed from the canonical Hamiltonian and the primary 
constraints of the theory. Therefore the Hamiltonian that 
generates evolution is
\begin{equation}
H =  {\cal S}_\lambda + {\cal V}_{\vec{\lambda}}
+ {\cal T}_\Lambda\,,
\label{eq:totalH}
\end{equation}
where
\begin{eqnarray}
\hspace{-9ex}
{\cal S}_\lambda &:=& \int_{\Sigma} \lambda \left[ G^{\mu \nu} 
\left( P_\mu-\frac{1}{\alpha}B_{\mu\sigma}\pi^\sigma \right)
\left( P_\nu-\frac{1}{\alpha}B_{\nu\rho}\pi^\rho \right)
 + \frac{1}{\alpha^2} \pi^A \pi^B h_{AB}  
+ \beta_p ^2 {\cal M} \right],
\label{eq:S}
\\
\hspace{-9ex}
{\cal V}_{\vec{\lambda}} &:=& \int_{\Sigma} \lambda^A  \left[
\left( P_\mu-\frac{1}{\alpha}B_{\mu\sigma}\pi^\sigma \right)
X^\mu_A + 
\frac{1}{\alpha} \pi^B {\cal F}_{AB} \right],
\label{eq:V}
\\
\hspace{-9ex}
{\cal T}_\Lambda &:=& \int_{\Sigma} \Lambda \,\pi^0\,.
\label{eq:O}
\end{eqnarray}
Here we have smeared out (\ref{eq:scalar}), (\ref{eq:vector}) and
(\ref{eq:gauss0}) by test fields $\lambda$, $\lambda^A$ and
$\Lambda$ in order to get phase space constraint functions. It is
worthy of notice that $\lambda$ is a test field with weight minus
one because the constraint (\ref{eq:scalar}) is of weight two.

Note that the appearance of the combination 
$Q_\mu:=P_\mu - \frac{1}{\alpha}B_{\mu\nu}\pi^\nu$ 
in both the scalar~(\ref{eq:scalar}) 
and vector~(\ref{eq:vector}) constraints plead to use a minimal prescription 
to incorporate the variable $Q$ as a natural momentum for the theory. 
However, a description of the Hamiltonian theory in terms of the momentum 
$Q$ makes unclear, to our purposes, the algebraic properties of the system 
since it brings along extra secondary constraints where terms proportional 
to the $B$ field and its derivatives emerge.  As we are interested in 
understanding the algebra of the constraints, as well as the Hamiltonian 
equations of motion, from now on we will only consider an 
everywhere vanishing $B$ field, and hence $Q$ is reduced to $P$.

It is worthy to mention that the geometrical nature of the momenta
$P$ help us to visualize another implicit scalar relation in the
set (\ref{eq:scalar})-(\ref{eq:gauss0}). From the definition of
$P$ note that 
\begin{equation}
\label{eq:Pdotpi}
 P \cdot \pi= 0\,,
\end{equation}
which encodes a privileged direction on the worldvolume where the 
dynamics can take place. This relation can lead to introduce another 
set of constraints in analogy to Virasoro constraints in string 
theory \cite{Lindstrom}.

\subsection{Poisson algebra}

Let $F$ and $G$ be two phase space functionals.  The Poisson
bracket of these two functionals is defined as
\begin{equation}
\{ F, G \}:= \int_{\Sigma}
\left(
\frac{\delta F}{\delta X}\cdot \frac{\delta G}{\delta P} +
\frac{\delta F}{\delta A_a} \frac{\delta G}{\delta \pi^a} -
\frac{\delta F}{\delta P} \cdot \frac{\delta G}{\delta X} -
\frac{\delta F}{\delta \pi^a} \frac{\delta G}{\delta A_a}
\right)\,.
\end{equation}
Hence the phase space is endowed with a symplectic structure that in terms of
its basis reads (at equal times)
\begin{eqnarray}
\{X^\mu(t,u),P_\nu(t,u')\} & = & \delta_\nu^\mu \delta^{(p)}(u,u')\,, 
\nonumber\\
\{A_a(t,u),\pi^b(t,u')\}   & = & \delta_a^b \delta^{(p)}(u,u')\,,
\label{eq:fundamentalPb}
\end{eqnarray}
and all other Poisson brackets are vanishing.

As usual, time evolution of any phase space functional $F$ can be
written as the Poisson bracket with the total Hamiltonian
(\ref{eq:totalH}), that is, $\dot{F} = \left\lbrace F, H
\right\rbrace $.  The constraints~(\ref{eq:S})-(\ref{eq:O}) must be 
preserved under the evolution of the system, and hence we impose 
the stability conditions
\begin{eqnarray}
\dot{{\cal S}}_\lambda & = & \left\lbrace {\cal S}_\lambda, 
H \right\rbrace \simeq 0\,,
\label{eq:dotS}\\
\dot{{\cal V}}_{\vec{\lambda}} & = & \left\lbrace {\cal V}_{\vec{\lambda}}, 
H \right\rbrace \simeq 0\,,
\label{eq:dotV}\\
\dot{{\cal T}}_\Lambda & = & \left\lbrace {\cal T}_\Lambda, 
H \right\rbrace \simeq 0\,.
\label{eq:dotT}
\end{eqnarray}
Conditions~(\ref{eq:dotS}) and~(\ref{eq:dotV}) are identically zero,
while the stability condition~(\ref{eq:dotT})
casts out the secondary constraint
\begin{equation}
  \varphi  = {\cal D}_A \pi^A \simeq 0,
\label{eq:gauss}
\end{equation}
which is the well known Gauss law associated to the $U(1)$
symmetry.
In a similar way as before, we can smear out the Gauss law with a test field
in order to get the phase space function
\begin{equation}
{\cal G}_\phi := \int_{\Sigma} \phi \,{\cal D}_A \pi^A\,.
\label{eq:Gauss}
\end{equation}
No further constraints are obtained if we impose
the stability condition on the constraint (\ref{eq:Gauss}).

The Poisson brackets of the constraints~(\ref{eq:S})-(\ref{eq:O})
with respect to~(\ref{eq:fundamentalPb}) satisfy the
algebra of the constraints,
\begin{eqnarray}
\{ {\cal V}_{\vec{\lambda}}, {\cal V}_{\vec{\lambda'}} \}&=&
{\cal V}_{[\vec{\lambda},\vec{\lambda'} ]} + {\cal G}_{\phi_1}\,,
\\
\{ {\cal V}_{\vec{\lambda}}, {\cal S}_\lambda \}&=&
{\cal S}_{{\cal L}_{\vec{\lambda}}\lambda} +
{\cal G}_{\phi_2}\,,
\label{eq:PB2}
\\
\{ {\cal S}_\lambda , {\cal S}_{\lambda'} \}&=&
{\cal V}_{\vec{\lambda^*}}\,,
\end{eqnarray}
where
\begin{eqnarray}
\phi_1 & := & \lambda^{'\,A}\lambda^B F_{AB}\,,\nonumber\\
\phi_2 & := & 2\lambda \lambda^A \pi^B h_{AB}\,,\nonumber\\
\lambda^{*\,A}
& := & 4\left[ \pi^A \pi^B + \beta^2 {\cal M} (M^{-1})^{(AB)}\right]
(\lambda {\cal D}_B \lambda' - \lambda' {\cal D}_B \lambda)\,.
\end{eqnarray}
Note that in equation (\ref{eq:PB2}) we have considered
that $\lambda$ is a scalar density of weight minus one, and hence
its Lie derivative along $\vec{\lambda}$ is given by ${\cal
L}_{\vec{\lambda}}\lambda = \lambda^A {\cal D}_A \lambda - \lambda
{\cal D}_A \lambda^A$ (see \ref{app:derivatives}). The
remaining Poisson brackets vanish strongly. Therefore we have a
first class constrained system that forms an open algebra since
the right hand side of the Poisson brackets involves structure
functions rather than structure constants.

An important feature of the Hamiltonian formalism is the exhibition
of the physical degrees of freedom transparently. In the DBI case the
counting of degrees of freedom will be as follows: $(1/2)[2(N+p+1)-2(p+3)]
= N-2$, corresponding to the $N+p+1$ total number of canonical variables 
and the $p+3$ number of first class constraints~(\ref{eq:scalar}-\ref{eq:gauss0}) and~(\ref{eq:gauss}).
Note then that the physical degrees of freedom are hence independent of
the dimension of the extended object. This fact was discussed
previously in \cite{Lee}.

Further, following the Dirac-Bergmann recipe for constrained systems, the
most important classification among the constraints in a physical
system is the one that separates them between first and second
class. Once we have identified the first-class constraints we are
able to write the extended first class Hamiltonian by
\begin{equation}
H_E = {\cal S}_\lambda + {\cal V}_{\vec{\lambda}}
+ {\cal T}_\Lambda + {\cal G}_\phi \,.
\label{eq:H}
\end{equation}
It is the extended Hamiltonian that provides the most general evolution
of the fields. In the next section we will study the Hamiltonian
equations of motion.

\section{Hamiltonian evolution equations}
\label{sec:Hamilton}

In this section we will check the way in which the Hamiltonian time evolution
relations reproduce the equations of motion obtained in 
Section~\ref{sec:DBIaction}.

We start by considering the time evolution of the $X$ coordinates
which reproduces the form of the
momentum $P$ given in (\ref{eq:P}),
\begin{equation}
\dot{X} = \left\lbrace X, H_E \right\rbrace =
\frac{\delta H_E}{\delta P} = 2\lambda P +
\lambda^A X{}_A\,,
\end{equation}
and also identifies the form of the Lagrange multipliers
$\lambda$ and $\lambda^A$.
So, contracting with the $\Sigma$ basis, we have the
expressions
\begin{eqnarray}
\lambda^A &=& N^A - \alpha\,(M^{-1})^{[AB]}{\cal F}_B \,,
\label{eq:lambdaA}
\\
\lambda &=&  \frac{\sqrt{-M}}{2\beta_p (-{\cal M})}\,.
\label{eq:lambda}
\end{eqnarray}
Once again, if we turn off the fields we are able to reproduce
the expressions reported in \cite{hambranes,ADMDNG}.

Similarly, time evolution of the $A_A$ coordinates identifies
the form of the momenta $\pi^A$ given in (\ref{eq:pi}),
\begin{eqnarray}
\dot{A}_A &=& \left\lbrace A_A,H_E \right\rbrace =
\frac{\delta H_E}{\delta \pi^A}= 2\lambda \pi_A +
\lambda^B F_{BA} - {\cal D}_A \phi \,,\nonumber \\
&=& N^B F_{BA} + {\cal F}_A - {\cal D}_A \phi \,,
\label{eq:ham2}
\end{eqnarray}
where we have introduced the Lagrange
multipliers~(\ref{eq:lambdaA}) and (\ref{eq:lambda}). In order 
to reproduce the equation (\ref{eq:A0}) we choose 
the value $\phi = -A_0$.

Now, the Hamiltonian equation for $A_0$ is
\begin{equation}
\dot{A}_0 = \left\lbrace A_0, H_E \right\rbrace =
\frac{\delta H_E}{\delta \pi^0} = \Lambda
\label{eq:A00}
\end{equation}
which shows us the explicit form for the Lagrange multipliers
necessary to recover the right equations of motion. Note that
(\ref{eq:ham2}) and (\ref{eq:A0}) are similar to the situation in
Maxwell theory due to the $U(1)$ symmetry. Thus,
equation~(\ref{eq:A00}) is only a pure gauge term.

Evolution of the momenta $\pi^A$ is given by
\begin{eqnarray}
\dot{\pi}^A &=& \left\lbrace \pi^A, H_E \right\rbrace =
- \frac{\delta H_E}{\delta A_A} \nonumber \\
&=&
\partial_B \left(
\alpha \sqrt{-{M}}\left( M^{-1}\right)^{[BA]} \right)
+ {\cal L}_{\vec{\lambda}}\pi^A     \nonumber\\
&=& \partial_B \left( N\sqrt{h}{\cal J}^{BA} \right)
+  {\cal L}_{\vec{\lambda}}\pi^A \,,
\end{eqnarray}
where in the second line of this equation we have inserted the Lagrange
multipliers~(\ref{eq:lambdaA}) and~(\ref{eq:lambda}), and we have 
introduced the spatial projection of the worldvolume bicurrent, that is, 
${\cal J}^{AB} = {\cal J}^{ab}X_a{}^AX_b{}^B$.

The Hamilton equation for the $\pi^0$ momentum reads
\begin{equation}
\dot{\pi}^0 = \left\lbrace \pi^0, H_E \right\rbrace =
- \frac{\delta H_E}{\delta A_0}=0\,,
\end{equation}
showing that $\pi^0$ is a constant of the motion, as expected.

Finally, time evolution for the momenta $P$ is given by
\begin{eqnarray}
\dot{P} &=& \left\lbrace P , H_E\right\rbrace =
- \frac{\delta H_E}{\delta X}\nonumber \\
&=& \partial_A \left[ 2\lambda \pi^A \pi^B\, X_{B}
- 2\lambda\,\alpha\,{\cal M}\,(M^{-1})^{(AB)}\,
X_{B} \right] + {\cal L}_{\vec{\lambda}}P  \nonumber \\
&=& \partial_A ( 2\lambda \pi^A \pi^B\, X_{B} )
- \partial_A (N\sqrt{h}\,T^{AB}\,X_{B}) + {\cal L}_{\vec{\lambda}}P \,,
\end{eqnarray}
where we have introduced the spatial projection of the
stress-energy tensor, $T^{AB} = T^{ab}X_a{}^A X_b{}^B$.

\section{$D1-$brane dynamics in $AdS_3\times S^3$}
\label{sec:kluson}

In order to illustrate the formalism developed previously, we consider
a $D1$-brane immersed in the background spacetime $ds^2= L^2\left[ 
-\cosh^2 \rho \,dt^2 + d\rho ^2 + \sinh^2 \rho \,d\theta_1 \right] +
L^2\left[ d\theta^2 + \cosh^2 \theta\,d\widetilde{\psi}^2 
+ \sin^2 \theta\,d\theta_2 ^2 \right] $ supported by the NS three form field
$H=dB = L^2 \sinh (2\rho)\,d\rho \wedge d\theta_1 \wedge dt$ where the
Kalb-Ramond two form field is $B=L^2 \sinh^2 \rho \, d\theta_1 \wedge dt$. We
assume that the worldsheet generated by the motion of the 
$D1$-brane is described by the embedding
\begin{equation}
 x^\mu = X^\mu (t,\theta_1) = (t,\rho(t),\theta_1,\theta, 
\widetilde{\psi},\theta_2)\,.
\label{eq:X}
\end{equation}
Further, we will assume that on the worldsheet lives the 
$U(1)$ gauge field $A_a = ( A_0 , A_1)=(0,A_{\theta_1})$ with 
$a,b=t,\theta_1$. Physically, this is a $D1$-brane with overcritical 
electric fields that wraps in $\theta_1$ direction, as discussed 
in~\cite{Bachas,KNP}. The components of the induced metric are given by
\begin{equation}
(g_{ab}) = L^2 \left( 
\begin{array}{cc}
-(\cosh^2 \rho - \dot{\rho}^2)& 0\\
0& \sinh^2 \rho
\end{array}
\right) \,.
\label{eq:indmetric}
\end{equation}
where we easily note that the square root of (minus) 
the determinant of the induced metric (\ref{eq:indmetric})
is given by $\sqrt{-g} = L^2 \sqrt{\cosh^2 \rho - \dot{\rho}^2}\,
\sinh \rho $. In a similar way, the non-null component of the matrix 
${\cal F}_{ab}$ is given by the expression ${\cal F}_{t\theta_1}:= 
\alpha  \dot{A}_{\theta_1} - L^2\sinh^2 \rho$.
The elements of the composite matrix are
\begin{equation}
 (M_{ab}) = \left( 
\begin{array}{cc}
-L^2 (\cosh^2 \rho - \dot{\rho}^2) &{\cal F}_{t\theta_1}\\
-{\cal F}_{t\theta_1}& L^2 \sinh^2 \rho  
\end{array}
\right) \,.
\label{eq:mab}
\end{equation}
The determinant of matrix~(\ref{eq:mab}) is hence given by $M = g + 
{\cal F}_{t\,\theta_1}^2$. The corresponding inverse matrix of (\ref{eq:mab}) 
is given by
\begin{equation}
 (M^{-1})^{ab} = \frac{1}{M} \left( 
\begin{array}{cc}
L^2 \sinh^2 \rho  &-{\cal F}_{t\theta_1}\\
{\cal F}_{t\theta_1}&- L^2 (\cosh^2 \rho - \dot{\rho}^2) 
\end{array}
\right).
\label{eq:mab-inv}
\end{equation}
It is straightforward to read both the expressions for the worldvolume
stress-energy tensor and covariant bicurrent,~(\ref{eq:Tab}) and~(\ref{eq:Jab}), 
respectively (see also equations~(\ref{eq:D1stress-energy}) 
and (\ref{eq:D1bicurrent})) 
\begin{eqnarray}
\sqrt{-g}\,T^{ab} &=& - \frac{\beta_1L^2}{\sqrt{-M}} \left( 
\begin{array}{cc}
 \sinh^2 \rho  &0\\
0 & - (\cosh^2 \rho - \dot{\rho}^2) 
\end{array}
\right),
\\
\sqrt{-g}\,J^{ab} &=& \frac{\alpha\beta_1\,{\cal F}_{t\theta_1}}
{\sqrt{-M}} 
\left( 
\begin{array}{cc}
0&-1\\
1 & 0 
\end{array}
\right).
\end{eqnarray}
The equation of motion (\ref{eq:eqmoto2}) (which is equivalent to 
equation~(\ref{eq:again})) can be written explicitly to show that 
$C_1:=\sqrt{-g}J^{\theta_1 t}$ equals a constant, which implies that
\begin{equation}
\left(\frac{{\cal F}_{t\theta_1}}{\sqrt{-g}}\right)^2 = 
\frac{C_1 ^2 }{\alpha^2 \beta_1 ^2 +
C_1 ^2}\,.
\label{eq:C1}
\end{equation}
Equation of motion~(\ref{eq:eqmoto1}) is complicated to solve since it 
involves a second-order nonlinear differential equation, however, in 
our treatment this equation is reduced to~(\ref{eq:D1extForce}) which 
can be easily manipulated to obtain a constant of motion associated to 
the energy of the $D1$-brane (times a $2\pi$ factor which comes from 
angular integration in $\theta_1$ direction)
\begin{equation}
\label{eq:constantenergy}
E := \frac{-G_{00}G_{22} \sqrt{C_1 ^2 + \alpha^2 \beta_1 ^2}-
C_1 B_{02} \sqrt{-g}}{\alpha \sqrt{-g}}\,.
\end{equation}
Equation~(\ref{eq:constantenergy}) is in agreement with the energy 
computed in \cite{KNP}, and is equivalent to the energy density 
$\varepsilon_1$ obtained above~(see~(\ref{eq:energy})). It is 
important to mention that this energy can also be obtained directly 
(up to a constant term) by taking the zero-th component of the momentum 
$P_\mu$ obtained in the ADM decomposition~(equation~(\ref{eq:P})).  
Note, however, that conservation of energy is guaranteed in this case 
due to the specific background we choose to work with, that is, a static 
background.  In a more general background, equation of motion~(\ref{eq:Moto1}) 
(or its reduced form~(\ref{eq:D1extForce})) follows instead. In addition, 
we can also compute the force density $F^{(1)}$ with the help of the 
constant of motion~(\ref{eq:C1}), and hence we find that the force density 
is also a constant of motion given by 
\be
F^{(1)}= -\frac{2C_1}{L\alpha} \,.
\ee 
Recall that we are interpreting this force density as an external force acting on 
the $D1$-brane through a generalized Newton's second law.

\section{Concluding remarks}
\label{sec:remarks}

In this paper we have carried out a geometrical study
of the classical DBI action. As a result of the variational process 
of the action we found out not only the equations of motion for a $Dp$-brane 
like a generalized Newton's second law but also a couple of conserved quantities 
which we identified as the stress-energy tensor and the worldvolume 
covariant bicurrent associated to the worldvolume electric induction. 
These equations of motion are written in a compact form. 
We found that these equations are exactly those of DNG theory
complemented with the ordinary inhomogeneous Maxwell theory, and it was shown that
our results are independent of the value of both, the tension of the
$Dp$-brane and the BI-parameter. 
Furthermore, our formulation allowed us to obtain the equations
of motion for systems with arbitrary background spacetimes. 
Additionally we note that for static backgrounds we obtained the conserved 
energy of the $Dp$-brane which is a very powerful result for exploring its dynamics.
We went through an specific example for a $D1$-brane in  
order to see how directly the results obtained by different approaches 
are reached by our geometric formulation.  We saw that in our formulation
the results are not only reproduced in an effortless way but also it allowed to
speculate with more general backgrounds. 
In~\ref{app:examples} we specialized our results 
to the simple cases of $D1$-branes and $D2$-branes. 
Further, we noted that the effect of the NS 2-form 
$B_{\mu \nu}$ field in both cases was to produce a conserved surface 
current. For higher $p$, we expect that a similar situation occurs for 
$Dp$-branes except that higher order invariants are involved in the 
equations of motion as compared
with the $D2$-brane case.  Indeed, this is the case for the $D3$-branes,
where a lengthy computation shows the emergence of the second invariant of 
the electromagnetic field that appears in the determinant of the composite 
matrix besides the inhomogeneous Maxwell equations. 
The geometrical study for the latter case will be considered in a separate work. 

On the other hand, we implemented a geometrical Hamiltonian analysis of the
DBI action based in the canonical ADM formalism of GR.
This was useful to break up the non-symmetric composite matrix $M_{ab}$
into its ``temporal" and ``spatial" parts.  We gained enough control
over the decomposition to recover the usual ADM decomposition of the
DNG extended objects by taking a vanishing BI-parameter.
Though the composite matrix $M_{ab}$ was written in terms of a
lapse function and a shift vector, we kept our original configuration
coordinates.  This allowed us to identify the phase space constraints in
a simple manner. We also discussed the algebra of the constraints of the theory.
As expected from experience in ADM general relativity, the algebra turned
out to be open with the structure functions given in a complicated way. 
This brings serious complications towards the study of the quantum theory of the DBI
action by means of canonical methods of quantization. 
Strictly 
speaking, until now is too difficult from first principles to fully quantise the 
theory,  but some
attempts are in progress \cite{Lindstrom, Usha, Kallosh}. 
We hope that our geometrical
approach pave the way to a consistent quantum analisys.
It will be interesting
to study convenient gauge conditions in order to partially overcome the
difficulties with the algebra of the constraints at both classical and quantum levels.
Once a gauge is specified, we expect to apply our general Hamiltonian 
approach in the research of the evolution of other geometrical interesting objects 
such as supertubes and superconducting tubes.
Also, it will
be important to study the physical observables of the theory once
an specific gauge is chosen.  This will be worked elsewhere.

\bigskip

\ack
We thank Plamen Bozhilov and David Fairlie 
for correspondence.
This work was supported in part by CONACYT (Mexico)
under grants C01-41639 and EP-050291. RC acknowledges partial support
from COFA and to the grants EDI and SIP 20061047. 
ER acknowledges partial support from PROMEP 2004-2007.

\appendix

\section{$D1$- and $D2$-brane general cases}
\label{app:examples}

It is instructive to consider the simplest cases in $Dp$-brane
theory to make the general development above transparent. Such 
simplest cases for which it is straightforward to explicitly
construct the stress-energy-tensor and the worldvolume bicurrent
are the $D1$-brane and the $D2$-brane systems.

\subsection{$D1$-branes}

In this case we have that the inverse matrix of the composite
matrix $M_{ab}$ as well as its determinant are
\begin{eqnarray}
\left( M^{-1}\right)^{ab} &=& \frac{g}{M} \left( g^{ab}
- {\cal F}^{ab}\right)\,,
\label{eq:inv1}
\\
M&=& g \left( 1 +  {\cal F}^2 \right)\,,
\end{eqnarray}
where we have introduced the notation ${\cal F}^2 := 
(1/2){\cal F}_{ab}{\cal F}^{ab}$. It is straightforward to read 
both the symmetric and antisymmetric parts of the composite matrix 
(\ref{eq:inv1}) and consequently the stress-energy tensor as well 
as the worldsheet covariant bicurrent are given by
\begin{eqnarray}
\label{eq:D1stress-energy}
T^{ab} &= &\frac{\beta_1}{\sqrt{1 +  {\cal F}^2}}\,g^{ab}\,,
\\
\label{eq:D1bicurrent}
{J}^{ab} &=& \frac{\alpha\beta_1}{\sqrt{1 +  {\cal F}^2}}\,
{\cal F}^{ab}\,.
\end{eqnarray}

We find that the equations of motion~(\ref{eq:eqmoto1}) and 
(\ref{eq:eqmoto2}) reduce to
\begin{eqnarray}
\label{eq:D1extForce}
K^i &=& \frac{\sqrt{1 + {\cal F}^2}}{\beta_1}F^i \,, \\
\label{eq:EMcurrent}
\nabla_a {F}^{ab}&=& {\cal J}_1 ^b \,,
\label{eq:again}
\end{eqnarray}
where we have used the relation $g^{ab}\nabla_a {\cal F}^2 =0$ from
(\ref{eq:eqmoto0}). 
Thus, in the simplest case, the resulting equations of motion 
seem similar to those of DNG theory with an external force complemented 
with the ordinary inhomogeneous Maxwell theory. 
Thus, the behaviour of the $U(1)$ gauge field is similar to that 
ocurring on the inhomogeneous Maxwell case with the current 
\begin{equation}
\label{eq:currentJ1}
{\cal J}_1 ^a := \alpha^{-1}\nabla_b B^{ab}\,.
\end{equation}
This current ${\cal J}_1 ^a$ is conserved, $\nabla_a 
{\cal J}_1 ^a=0$. 
Note that the effects of the NS field are to produce both a conserved 
surface current on the worldvolume and a force density $F^i$.

\subsection{$D2$-branes}

Now we specialize to the case of a $D2$-brane.
This case is slightly more complicated. The inverse matrix as 
well as the determinant of the composite matrix~$M_{ab}$ are
\begin{eqnarray}
\left( M^{-1}\right)^{ab} &=& \frac{g}{M}\left\lbrace \left[  \left( 1 +
 {\cal F}^2 \right)g^{ab} +  {\cal F}^{ac}{\cal F}_c{}^b  
\right] -  {\cal F}^{ab}\right\rbrace \,,
\label{eq:inv2}
\\
M &=& g \left( 1 +  {\cal F}^2 \right)\,.
\end{eqnarray}
As for the $D1$-branes, it is straightforward to read both the
symmetric and antisymmetric parts of the composite matrix
(\ref{eq:inv2}), and we can identify both conserved stress-energy
tensor and the physical tensor ${J}^{ab}$,
\begin{eqnarray}
T^{ab} &=& \frac{\beta_2}{\sqrt{1 +  {\cal F}^2 }}\,
\left[ (1+ {\cal F}^2) g^{ab} +  {\cal F}^{ac}
{\cal F}_c{}^b \right] \,,
\\
{J}^{ab} &=& \frac{\alpha\beta_2}{\sqrt{1 +  {\cal F}^2 }}\,
{\cal F}^{ab}\,.
\end{eqnarray}
On one side, we find that for this case the equations of 
motion~(\ref{eq:eqmoto1}) and~(\ref{eq:eqmoto2}) become
\begin{eqnarray}
K^i & = & \frac{1}{\beta_2\sqrt{1+ {\cal F}^2}}( F^i - {\cal F}^i)\,, 
\label{eq:D2-Newton}
\\
\nabla_a F^{ab}      & = & {\cal J}_2 ^b         \,,
\label{eq:D2-IMaxwell}
\end{eqnarray}
where we have defined ${\cal F}^i := \beta_2 (1+{\cal F}^2)^{-1/2}
 {\cal F}^a{}_b{\cal F}^b{}_c K_a{}^{c\,i}$, and the current density vector 
\begin{equation}
\label{eq:currentJ2}
{\cal J}_2 ^b  = 
\frac{(\nabla_a {\cal F}^2)
\,{\cal F}^{ab}} {2\alpha (1+{\cal F}^2)} + {\cal J}_1^b\,,
\end{equation}
where ${\cal J}_1^b$ is defined analogously to~(\ref{eq:currentJ1}) 
for the case of $D2$-branes.  The current~(\ref{eq:currentJ2}) 
is also conserved, and hence we have $\nabla_a {\cal J}_2 ^a = 0$. 
Further, equation~(\ref{eq:D2-IMaxwell}) represents the inhomogeneous 
Maxwell equations supported on the worldvolume.

On the other side, we see that the conservation law~(\ref{eq:eqmoto0})
can be written as
\be
\label{eq:Maxwellconservation}
\nabla_a \Theta^{ab}  =  -{\cal F}^{ba}({\cal J}_{2\,a} -
{\cal J}_{1\,a})\,,
\ee
by defining the symmetric stress tensor
$\Theta^{ab}:=\frac{1}{2} g^{ab} {\cal F}^2 + {\cal F}^{ac}
{\cal F}_c{}^b$. This equation for $\Theta^{ab}$ stands for the 
analogous of the conservation of energy and momentum equations 
for electromagnetic fields interacting with sources described by 
the current ${\cal J}^a:={\cal J}^a _{2} - {\cal J}^a _1$ in 
the Maxwell case. In fact, the equation
(\ref{eq:Maxwellconservation}) is expected due to the presence 
of the external NS field. It is worth noticing that, contrary to the
Maxwell case, the tensor $\Theta^{ab}$ has non-vanishing trace.
This is a peculiarity of the non-linear electrodynamics. 
Further, the term ${\cal F}^{ba}({\cal J}_{2\,a} - {\cal J}_{1\,a})$ 
appearing in equation~(\ref{eq:Maxwellconservation}) can be recognised 
as a Lorentz force density~\cite{Jackson}.

\section{Important mathematical identities}
\label{app:mathidentities}

\subsection{Determinant of a matrix}
\label{app:det}

With the Levi-Civita tensor, $\epsilon^{a_1 a_2 \cdots a_n}$, in
$n$-dimensions, we can build many invariant quantities by means of
its properties as well as provide an elegant and compact way to
obtain many of the relevant relations in matrix algebra. The
Levi-Civita tensor is related to the totally antisymmetric
pseudotensor by the relation
\begin{equation}
\epsilon_{a_1 a_2 \cdots a_n} = \sqrt{-g} \,\varepsilon_{a_1 a_2
\cdots a_n}\,,
\end{equation}
where $\varepsilon_{a_1 a_2 \cdots a_n}$ is the Levi-Civita
pseudotensor which is a tensorial density of weight $w = -1$.

The determinant of a matrix $M_{ab}$ can be defined in terms of
the Levi-Civita pseudotensor by
\begin{equation}
M:= {\mbox{det}}\,(M_{ab})=\frac{1}{n!}\varepsilon^{a_1 a_2 \cdots
a_n} \varepsilon^{b_1 b_2 \cdots b_n} M_{a_1 b_1}M_{a_2 b_2}
\cdots M_{a_n b_n}\,,
\end{equation}
where we assume that $\varepsilon_{12\cdots n}= 1$.

The inverse matrix, for the case when $M_{ab}$ is nonsingular,
also has a representation by means of the Levi-Civita pseudotensor
\begin{equation}
(M^{-1})^{a_1b_1}= \frac{1}{(n-1)!\,M}\,\varepsilon^{a_1 a_2
a_3\cdots a_n} \varepsilon^{b_1 b_2 b_3\cdots b_n}
M_{b_2a_2}M_{b_3a_3} \cdots M_{b_n a_n}\,.
\end{equation}

\subsection{Geometrical identities involving $M_{ab}$}
\label{app:identities}

There are powerful identities involving the inverse of the matrix
$M_{ab}$ and its worldvolume components. These are given by
\begin{eqnarray}
(M^{-1})^{(ac)}g_{cb} +  (M^{-1})^{[ac]}{\cal F}_{cb} 
&=& \delta^a {}_b
\label{eq:ident1}
 \\
(M^{-1})^{[ac]}g_{cb} +  (M^{-1})^{(ac)}{\cal F}_{cb}
&=& 0.
\label{eq:ident2}
\end{eqnarray}
Exist similar identities for the spatial contraparts $(M^{-1})^{AB}$,
$h_{AB}$ and ${\cal F}_{AB}$.

\subsection{Derivatives of tensor densities}
\label{app:derivatives}

The Lie derivative of an arbitrary tensor density of
weight $n$ along the direction of the field $\lambda^A$ is given by
\begin{eqnarray}
\fl
{\cal L}_{\vec{\lambda}}T_{A_1 A_2 \ldots A_s}{}^{B_1 B_2
\ldots B_r}
&=& \lambda^C\partial_C T_{A_1 A_2 \ldots A_s}{}^{B_1 B_2
\ldots B_r} + (\partial_{A_1}\lambda^C)T_{C A_2 \ldots A_s}
{}^{B_1 B_2 \ldots B_r} + \ldots \nonumber \\
&-&  (\partial_{C}\lambda^{B_1})  T_{A_1 A_2 \ldots A_s}
{}^{C B_2 \ldots B_r}
+ n (\partial_{C}\lambda^C) T_{A_1 A_2 \ldots A_s}{}^{B_1
B_2 \ldots B_r}.
\end{eqnarray}
In a similar way, the covariant derivative of an arbitrary
tensor density of weight $n$, is given by
\begin{eqnarray}
\fl
{\cal D}_C T_{A_1 A_2 \ldots A_s}{}^{B_1 B_2 \ldots B_r} &=&
\partial_C T_{A_1 A_2 \ldots A_s}{}^{B_1 B_2 \ldots B_r} +
\Gamma^{B_1} _{CD} T_{A_1 A_2 \ldots A_s}{}^{D B_2 \ldots B_r}
+ \ldots \nonumber \\
&-& \Gamma^{D} _{A_1 C} T_{D A_2 \ldots A_s}{}^{B_1 B_2
\ldots B_r} - \ldots - n  \Gamma^{D} _{ C D}  T_{A_1 A_2
\ldots A_s}{}^{B_1 B_2 \ldots B_r}.
\end{eqnarray}

\end{document}